\documentclass[structabstract]{anonymous}  
\usepackage{graphicx}
\usepackage{natbib}
\usepackage{txfonts}
\begin{document}
   \title{The white-light continuum in the \\impulsive phase of a solar flare}


   \author{H.~S.~Hudson
         \inst{1}, L.~Fletcher
         \inst{2}, and S.~Krucker\inst{1}
          }
           
   \institute{$^1$Space Sciences Laboratory, UC Berkeley,
              7~Gauss Way, Berkeley CA USA 94720-7450\\
              $^2$University of Glasgow, UK}
                

 \abstract{
We discuss the IR/visible/VUV continuum emission of the impulsive phase of a solar flare, using TRACE UV~and
EUV~images to characterize the spectral energy distribution.
This continuum has been poorly observed but dominates the radiant energy energetically.
Recent bolometric observations of solar flares furthermore point to the impulsive phase as the
source of a major fraction of the radiant energy.
This component appears to exhibit a Balmer jump and thus must originate in an optically thin
region above the quiet photosphere, with an elevated temperature and strong ionization.
}  

\titlerunning{White-light flares}

\maketitle
   \date{}
   
\section{Introduction}

Since the original observation \citep{1859MNRAS..20...13C} it could be recognized that the visible continuum 
might represent intense energy release -- the flare emission rivaled that of the photosphere itself.
Until recently it has been difficult to quantify this energy \citep{2007ApJ...656.1187F}, since the extension of the white-light continuum into the ultraviolet wavelengths could not readily be observed.
To this date we have only an extraordinarily skimpy record of the UV~and VUV extension of 
the white-light continuum.
White-light spectra in visible light also remain rare, principally because of the difficulty of placing
a spectrograph slit at the right location and at the right time.
The flare continuum, though bright, must compete with spatial and temporal ``noise'' from the photosphere.
This amounts to contrast fluctuations ($\Delta I/I$) of a few percent even in the quiet Sun, due mainly to convective motions (granulation) and to the p-modes.
The spatial variations are much larger in active regions where flares tend to occur.

In the meanwhile flare observations at other wavelengths have proliferated. 
The X-ray and radio observations showed that the upper solar atmosphere, into the corona, must play a vital role in flare physics.
Observations in the UV~beyond the Balmer limit ($\lambda < 3646$\AA) and the vacuum~UV (VUV) up to Ly-$\alpha$ have lagged, even though such observations would clarify the energy content implied by the visible continuum.
Most recently \citet{2004GeoRL..3110802W} have for the first time detected solar flares bolometrically
\citep[see also][]{1983SoPh...86..123H}.
Other non-visible observations have provided a great deal of diagnostic information.
In particular a strong association between the occurrence of the white-light coninuum and the hard X-ray emission was noted early on \citep{1970SoPh...13..471S,1975SoPh...40..141R}
and convincingly established with modern data \citep{1992PASJ...44L..77H,1993SoPh..143..201N}.

Optical spectra of white-light flares have been divided into type~I and type~II acoording to the presence  or absence of the Balmer jump in the optical continuum \citep{1986lasf.conf..483M}.
In this paper we argue that the type~I events, which show the Balmer jump, are associated with the impulsive phase of the flare and consist of continuum produced in an optically-thin layer with non-equilibrium ionization at some altitude, presently not known, above the quiet photosphere.
\citet{1985SoPh...98..255B} describe the interpretation of such a slab model, but without the extra ionization.

\section{The white-light spectrum}

At visible wavelengths, the white-light continuum in the quiet Sun comes from H$^-$~free-free and free-bound radiation.
Note that the H$^-$~free-free process does not actually involve an H$^-$~ion as such.
At longer wavelengths ordinary free-free opacity due to electron-proton interactions begins to dominate.
During flares this situation may change because of increased ionization.
Because the radiation helps to define the actual structure of the atmosphere, we must simultaneously understand the plasma motions and the radiative transfer processes.
Especially during the impulsive phase we cannot assume that the continuum would have the same sources as during quiet times, for example as described in semi-empirical static models such as VAL-C \citep{1981ApJS...45..635V}.

\begin{figure}
   \centering
   \includegraphics[width=6cm]{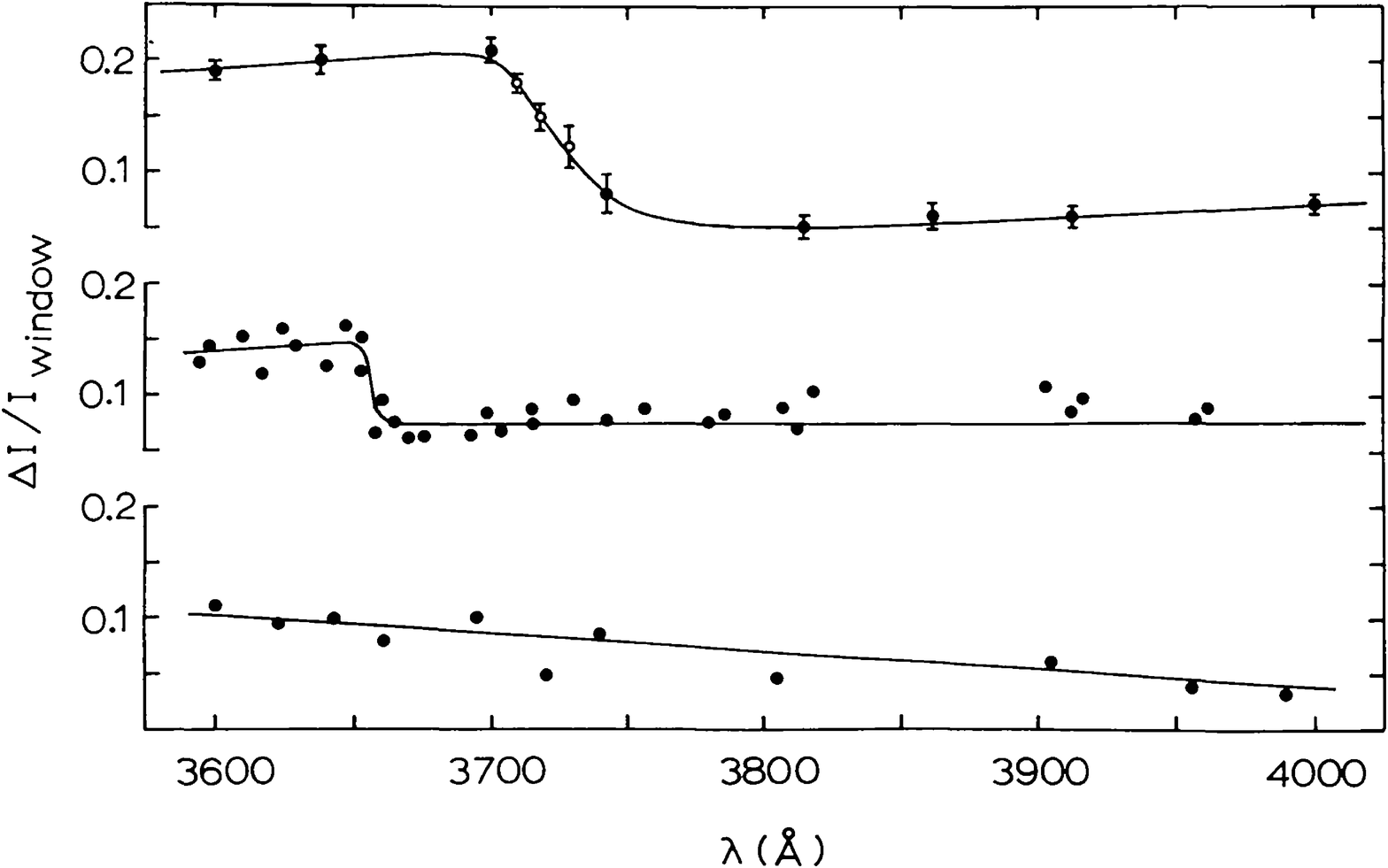}
      \caption{Comparison of UV~spectra of three white-light flares, from Neidig (1983): (\textit{upper}), Hiei (1982, \textit{middle}), and Machado~\& Rust (1974, \textit{lower}).
              }
         \label{fig:neidig}
   \end{figure}

Observations of the flare white-light continuum itself have have proven difficult, in spite of its importance.
Few dedicated observations explicitly dedicated to studying these phenomena could be properly motivated, given the rarity of the phenomenon.
D.~Neidig of Sacramento Peak Observatory led the most important one, which included broad-band photomeric observations, spectroscopic observations, and theoretical interpretation.
We reproduce a figure from \citep{1984SoPh...92..217N} in Figure~\ref{fig:neidig} which illustrates some of the 
interesting problems of white-light flare (WLF) spectroscopy.
in particular one can note that the continuum jump expected at the Balmer edge (3646\AA)~appears clearly in the \citet{1982SoPh...80..113H} spectrum, not at all in the \citet{1974SoPh...38..499M} spectrum, and apparently in the \textit{wrong place} in the 
\citet{1983SoPh...85..285N} spectrum.
This latter peculiarity -- observed as well in other flares -- could conceivably be explainable as the result of the crowding-together of higher Balmer lines, which could make the identification of the continuum ambiguous, but \citet{1981ApJ...248L..45Z} argue against this suggestion; the origin of this ``blue continuum'' therefore remains somewhat ambiguous.

\section{Flare continuum mechanisms}

A flare alters the structure of the solar atmosphere, affecting all of the plasma parameters; the complexity of the radiation processes and the fact that they are intimately coupled to the structure itself makes this a complicated theoretical problem.
Accordingly many authors have sought understanding through  numerical calculations in the radiation hydrodynamics in the 1D radiation hydrodynamics approximation.
Such a complicated approach is necessary even for the relatively simple hydrogen atom, since it dominates the continuum and since the particular continuum components (Lyman, Balmer, Paschen) depend on the non-equilibrium level populations of many levels.
Essentially the problem of continuum formation boils down to an understanding of the variation of opacity and electron temperature as a function of height and time in the perturbed atmosphere; T$_e$ at a range of optical depths determines the surface brightness. 
Unfortunately 3D~structure may be involved \citep[e.g.][]{2007PASJ...59S.807I}, and so even the concept of ``height'' needs to be examined critically.

For thermal processes, we can use the basic principles of radiative transfer to describe a slab of warm chromospheric material overlying the unperturbed photosphere.
For slab optical depth $\tau$, the emitted flux 
$$F \propto  (1 - e^{-\tau})\  {\mathrm S}_{slab} + {\mathrm S}_{phot}\ e^{-\tau},$$
where S is the source function.

Kirchoff's Law states that  $j_{\nu}\ =\  \kappa_{\nu} S_{\nu}$, where the source function S$_\nu$ would be the Planck function if the medium were in local thermodynamic equilibrium.
Thus  to produce an excess emission from the chromosphere, both the ionization (to enhance the continuum)
and the temperature (to increase the source function) must increase during the flare \citep{1972SoPh...24..414H}.
This can result from collisional heating and the associated non-thermal ionization mainly from the secondary effects of  ``knock-on'' electrons.
Each primary ionization event results in an almost equally fast electron.
In the ``specific ionization''  approximation to the ionization rate,  one ion pair is formed for approximately each 36~eV of energy deposition for neutral hydrogen \citep{1958RSPSA.248..415D}.
More detailed calculations confirm the basic importance of non-thermal ionization in this context \citep{1987A&A...174..270A}
The secondary ionization may non-local, so the physics in this approximation could similar to that of ``back-warming,'' where in the Balmer continuum photons can penetrate deeper than the layer of primary energy deposition, and produce not only secondary ionization but also heating in a deeper layer.

We do not think that a self-consistent theory including radiative transfer together with a detailed description of the electron distribution function yet exists.
The ionization in the solar atmosphere in general does not obey the Saha equation, since the medium is optically thin.
Accordingly one must consider not only non-thermal ionization, but also non-equiliibrium effects \citep{1991SoPh..135...65H} resulting from the asymmetry of ionization and recombination rates.
These effects (non-thermal and non-equilibrium) imply that the ionization of hydrogen can deviate from that expected in any calculation of ionization equilibrium, and of course these deviations should be most pronounced during the sudden perturbations associated with a flare.
The significance of overionization for continuum emission can readily be seen from the emission measure: hydrogen continuum depends basically on n$_e^2$, whereas any line emission varies as n$_e$n$_Z$, where n$_Z$ is the particular ionic species involved.
Overionization produces an excess of electron pressure and thereby increases the continuum/line ratio in regions with intermediate hydrogen ionization.
The detection of the Paschen continuum (from 8204\AA)  by \citet{1984SoPh...92..217N} provides evidence that an optically thin layer could generate the continua observed all across the IR/UV/VUV range.
Note that \citet{2001ApJ...563L.169L} also observed Brackett continuum excesses near the Ca~{\sc ii} line at 8542\AA~in a white-light flare. 

\section{Observations}

Only a few solar observations of the UV/EUV spectra of solar flares exist.
Notably \citet{1996ApJ...468..418B} observed an X-class flare, but temporal information was almost completely missing.
There are however many broad-band imaging observations of flares, especially from TRACE.
The EUV passbands (303\AA, 284\AA, 195\AA, and 171\AA) are normally interpreted in terms of their
dominant line contributions, for example as obtained from CHIANTI.
\citet{1999ApJ...511L..61F} point out that these passbands also respond to the continuum.
In this analysis we assume that strong nonthermal ionization makes the continuum dominate these passbands and use them to sketch out the spectral energy distribution.

\begin{figure}[h]
   \centering
  \includegraphics[width=6.5cm]{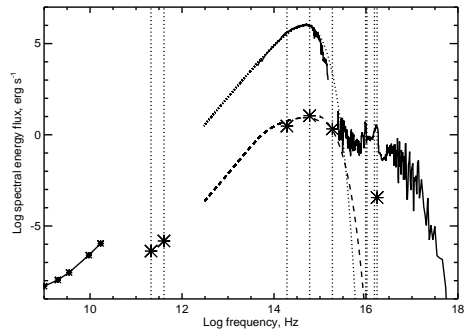}
      \caption{The spectral energy distribution ($\nu$f$_\nu$) of the quiet Sun, with an optically thin 
      warm chromospheric slab source superposed on the photosphere and shown as a dashed line.
      The parameters for this spectrum are 1.4~$\times$~10$^4$~K, 10$^{14}$~cm$^{-3}$, emission measure 
      3~$\times$~10$^{54}$~cm$^{-3}$, and line-of-sight thickness 10$^7$~cm.
      The visible/IR quiet-Sun data come from Wehrli (1985), the UV beyond 1200\AA~from Warren (2005), and the microwave points from the Nobeyama fixed-frequency data for December 2008.
      The dotted line shows a blackbody at 5778~K.
      The flare points (asterisks) show data from \citet{2004ApJ...603L.121K}, \citet{2004ApJ...607L.131X}, and our analysis of TRACE observations at 1600\AA~and 171\AA, with all points scaled by the GOES magnitude of the event to a common X10 level.
              }
         \label{fig:sacplot_thin}
   \end{figure}

Figure~\ref{fig:sacplot_thin} shows TRACE observations (1600\AA~and 171\AA) of a C6.7~flare (2003 Oct.~20) normalized to GOES~X10 and compared with the TSI observation of Woods et al. (2004).
This composite, represented as $\nu f_\nu$ to show the energy distribution, is consistent with the conclusion that the bulk of the energy radiated in the impulsive phase of a flare lies in the visible or near~UV spectrum.
This result agrees with that of \citet{2007ApJ...656.1187F}.

An optically thin slab lying above the photosphere will show recombination edges and a (hydrogen) free-free flux given by $S_\nu = \kappa_\nu B_\nu L$, or
$$
S_\nu \approxeq 0.1 {{n_e^2}\over{\nu^2 T^{3/2}}} B_\nu {{A\Delta h}\over{4\pi R_\odot^2}} \ \  \  {\rm erg (cm^2 sec)^{-1}}
$$
where the flare area-height product $A\Delta h$ can be determined from the (volumetric) emission measure and the density.
Here we assume that hydrogen is fully ionized, the free-free Gaunt factor is unity, and $Z_\odot = 1.2$ using
the formulae of \citet{1987A&A...183..341K}.
The energy distribution in Figure~\ref{fig:sacplot_thin} shows an example of this, assuming a large density and a small line-of-sight thickness in the spirit of the \citet{1977PAZh....3..315S} ``constant density'' approximation of the ambient atmosphere.
In this approximation the energy deposition occurs more rapidly than hydrodynamic effects can change the preflare density structure appreciably.

The example shown in Figure~\ref{fig:sacplot_thin} shows what should happen if a relatively deep layer
(the density corresponsd to a height above the photosphere of about 800~km in the VAL-C model) becomes fully ionized.
This layer is at a column depth of about 10$^{21}$~cm$^2$, and so corresponds to an electron energy of about  100~keV in the \citet{1971SoPh...18..489B} propagation model.
This is  roughly consistent with the energetics for white-light flares \citep{2007ApJ...656.1187F}.
but it is definitely inconsistent with the heating requirements for white-light flare models requiring the source to be at the pre-flare photosphere in height.
 
In both the optically-thick and optically-thin interpretations of the visible/UV continuum, the observations show that the bulk of the energy radiated in the impulsive phase lies at longer wavelengths; the EUV and X-ray domains contribute relatively little.
Furthermore, the required source densities, consistent with the observed source areas \citep[e.g.,][]{2006SoPh..234...79H} imply heights deep in the solar atmosphere but still many scale heights above the visible pre-flare photosphere.

The emission measure required for an optically-thin source at visible wavelengths is 3-4 orders of magnitude larger than that observed in soft X-rays or in the microwave ``post-burst increase'' sources \citep{1972SoPh...23..155H,1995SoPh..160..181C}.
This implies that the source must be compact enough to be optically thick at centimeter wavelengths.
We check that this is the case by computing the frequency for which the optical depth $\tau_\nu$~=~1.
For the model parameters of Figure~\ref{fig:sacplot_thin} this frequency turns out to be well above 1~THz (in fact, at about 10~$\mu$), consistent with the THz emission component observed by \citet{2004ApJ...603L.121K} and by \citet{2004A&A...415.1123L}.
We note that the warm chromospheric slab source cannot explain the EUV observations without requiring too much energy.

\section{Conclusions}

We have made the case that the impulsive-phase continuum emission at IR, visible, and near UV wavelengths results from an optically thin source layer well above the nominal photosphere in spite of the 1.56$\mu$~observations of \citet{2004ApJ...607L.131X}.
This layer, to be bright in the continuum, must be over-ionized as a result of non-thermal and non-equilibrium processes that remain to be described properly.
The observed strong association between white-light flaring and hard X-ray emission implies that non-thermal electrons play a dominant role in energy transport, and also control the ionization.
Whether these electrons come from a coronal acceleration ``black box'' as in the standard thick-target model 
\citep{1971ApJ...164..151K,1971SoPh...18..489B,1972SoPh...24..414H} remains to be seen, since other mechanisms of energy transport may be required \citep{2008ApJ...675.1645F,1982SoPh...80...99E}.

The model comparison we have shown in Figure~\ref{fig:sacplot_thin} should not be overinterpreted, since we had to intercompare data from three different flares.
The plot also has a range of many decades, which tends to hide discrepancies, and even a simple slab model has four free parameters (T, n$_e$, $\Delta$h, and the emission measure).
The most discrepant parameter in the model of Figure~\ref{fig:sacplot_thin} would be the derived horizontal scale $\sqrt{A}$ = 1.7~$\times$~10$^9$~cm.
We hope in the future to examine more of the EUV and X-ray data with the same continuum interpretation we have used here; note that Feldman et al. (1999) found that even in the post-maximum phase of their C6.8 flare, continuum emission still appeared to dominate the EUV passbands of TRACE.

The extension of the white-light flare spectrum into the VUV, and possibly beyond, remains as a goal for future observations.
We do not know what spectra can match the TSI observations of Woods et al.  (2004; M. ~Kretzschmar, personal communication 2008), in terms of mean parameters or detailed atmospheric structures.
Because a flare perturbs the pre-flare atmosphere so dramatically, involving not only non-equilibrium and non-thermal atomic processes but also strong mass motions from unknown causes on short (and probably unresolved) time scales, we believe that modeling presents difficult problems that have not yet been tackled.
Thus continuum observations across the entire range from the far IR~through to the EUV will provide our best understanding of the dominant radiation processes of a flare.

\begin{acknowledgements}
 The authors acknowledge support from NASA via contract NAS 5-98033 for RHESSI, from STRC via rolling grant ST/F002637/1, and from the EU's SOLAIRE Research and Training Network (MTRN-CT-2006-035484).
We especially thank the International Space Science Institute (Bern, Switzerland) for support during the preparation of this manuscript.
Author Fletcher also acknowledges travel support from the National Solar Observatory (Kitt Peak).
\end{acknowledgements}

\bibliographystyle{aa}
\bibliography{hudson}

\end{document}